\documentclass[english,letterpaper,aps,prl,twocolumn]{revtex4}
\usepackage{lmodern}
\usepackage[T1]{fontenc}
\usepackage[latin1]{inputenc}
\usepackage{color}
\usepackage{amsmath}
\usepackage{graphicx}
\usepackage[normalem]{ulem}

\makeatletter

\providecommand{\tabularnewline}{\\}

\@ifundefined{textcolor}{}
{%
 \definecolor{BLACK}{gray}{0}
 \definecolor{WHITE}{gray}{1}
 \definecolor{RED}{rgb}{1,0,0}
 \definecolor{GREEN}{rgb}{0,1,0}
 \definecolor{BLUE}{rgb}{0,0,1}
 \definecolor{CYAN}{cmyk}{1,0,0,0}
 \definecolor{MAGENTA}{cmyk}{0,1,0,0}
 \definecolor{YELLOW}{cmyk}{0,0,1,0}
}

\makeatletter

\usepackage{epsfig}
\usepackage{dcolumn}
\usepackage{bm}
\usepackage{bbm}
\usepackage{wasysym}
\usepackage{marvosym}

\usepackage{subfloat}
\usepackage{color}
\usepackage{subfigure}

\usepackage{blindtext}

\newlength{\textwidthm}

\usepackage{tikz}

\makeatother

\usepackage{babel}
\usepackage{pgffor}
\usepackage{pdfpages}

\begin{document}

\title{Ferromagnetic Mott State in Twisted Graphene Bilayers at the Magic
Angle}

\author{Kangjun Seo$^{1}$, Valeri N. Kotov$^{2}$, and Bruno Uchoa$^{1}$$^{*}$ }

\affiliation{$^{1}$Department of Physics and Astronomy, University of Oklahoma,
Norman, OK 73069, USA}
\email{uchoa@ou.edu}
\affiliation{$^{2}$Department of Physics, University of Vermont, Burlington, VT 05405, USA}

\date{\today}
\begin{abstract}
We address the effective tight-binding Hamiltonian that describes
the insulating Mott state of twisted graphene bilayers at a magic angle. In that
configuration, twisted bilayers form a honeycomb superlattice of localized
states, characterized by the appearance of flat bands with four-fold
degeneracy. After calculating   the maximally localized   superlattice Wannier
wavefunctions, we derive the effective spin model
that describes the Mott state. We suggest that
 the system is an exotic \emph{ferromagnetic} Mott
insulator, with well defined experimental signatures. 
\end{abstract}
\maketitle
\emph{Introduction.$-$ }Mott insulators describe materials that exhibit
insulating behavior as a result of strong local interactions \cite{imada}.
In those systems, strong on site repulsion penalizes the kinetic energy
for electrons to hop between sites, rendering the electronic orbitals
localized. The strong degree of localization of the electronic wavefunction
favors antiferromagnetic alignment of the spins due to Pauli principle \cite{Auerbach}.  
Recent experiments \cite{Cao1,Cao2} indicate that
twisted graphene bilayers have a Mott state with an activation gap
of $\Delta\approx0.3$ meV that undergoes a metal-insulator transition
in the vicinity of a superconducting phase \cite{Cao2, Yankowitz}. This system is purely made of carbon atoms, with additional degrees
of freedom inherited from graphene \cite{Castro Neto}.  That has motivated
the question of whether the observed state could
be described by a novel Mott insulator \cite{Po2} or other exotic
correlated states \cite{Padhi,Xu,Irkhin,Thomson,dodaro}. Unveiling the nature of the insulating state
may be key to explain some of the the remarkable properties in the metallic phase. 

By twisting two graphene sheets at a small angle of the order of $\theta\sim1.1^{\circ}$,
what was dubbed a ``magic'' angle, interference due to hopping between
the layers leads to a Moire pattern and to a significant reconstruction
of the mini bands in the Moire Brillouin zone, which become flat \cite{santos,Bistritzer }.
Those flat bands have four-fold degeneracy, which is reminiscent of
the valley and spin quantum numbers of the graphene sheets. In general,
the confinement of interacting Dirac fermions in flat bands is expected
to create an emergent SU(4) symmetry, as previously predicted in graphene
heterostructures \cite{Uchoa1,Dou, Xu3} and in graphene Landau
levels \cite{Goerbig,Young0,Young,Nomura,Alicea,Abanin,Sodemann}.
Here, the Moire pattern forms a superlattice of quasi-localized states
with the size of the unit cell set by the twist angle, as shown in
Fig. 1. 

In this Letter, we show that the low energy Hamiltonian of the flat
bands at quarter filling maps into the \emph{ferromagnetic} spin exchange
Hamiltonian on a honeycomb superlattice,
\begin{equation}
\mathcal{H}=-\sum_{ij}J_{ij}\!\left(\frac{1}{2}+2\tau_{i}^{z}\tau_{j}^{z}-2\eta_{ij}\boldsymbol{\tau}_{i}^{\perp}\cdot\boldsymbol{\tau}_{j}^{\perp}\right)\!\left(\frac{1}{2}+2\mathbf{S}_{i}\cdot\mathbf{S}_{j}\right)\!,\label{eq:H1}
\end{equation}
where $\mathbf{S}_{i}$ is the localized spin on a superlattice site
$i$, $\boldsymbol{\tau}_{i}=(\tau^{x},\tau^{y},\tau^{z})\equiv(\boldsymbol{\tau}^{\perp},\tau^{z})$
is an orbital pseudospin operator that is reminiscent of the valley
quantum numbers, and $J_{ij}>0$ is the exchange coupling. The parameter
$\eta_{ij}=-1$ when $i,j$ belong to the same sublattice, in which
case the exchange interaction has SU(4) symmetry, and $\eta_{ij}=1$
otherwise, including  nearest neighbor (NN) sites. 
This Hamiltonian acts in the Hilbert space which is spanned
by four degenerate states per site, $|\alpha,\sigma\rangle$, with
$\alpha=\pm$ and $\sigma=\uparrow,\downarrow$ for the two orbital
pseudospins and spin quantum numbers respectively. 

The existence of direct exchange ferromagnetism in an insulating state
is uncommon \cite{Uchoa1} and reflects the very unusual shape of
the Wannier orbitals in this system. Ferromagnetism has been recently
observed in insulating van der Waals heterostructures of magnetic
chromium trihalide materials, CrX$_{3}$ (X$=$I, Br, Cl) \cite{Huang,Yelon,Samuelson},
which have crystalline field anisotropies that produce an ordered
Ising state. To the best of our knowledge, we are not aware of any
examples of ferromagnetic Mott states which do not involve orbital
ordering via a superexchange mechanism \cite{Erickson,Kugel2}.

\begin{figure}[b]
\vspace{-0.3cm}\includegraphics[width=0.98\linewidth]{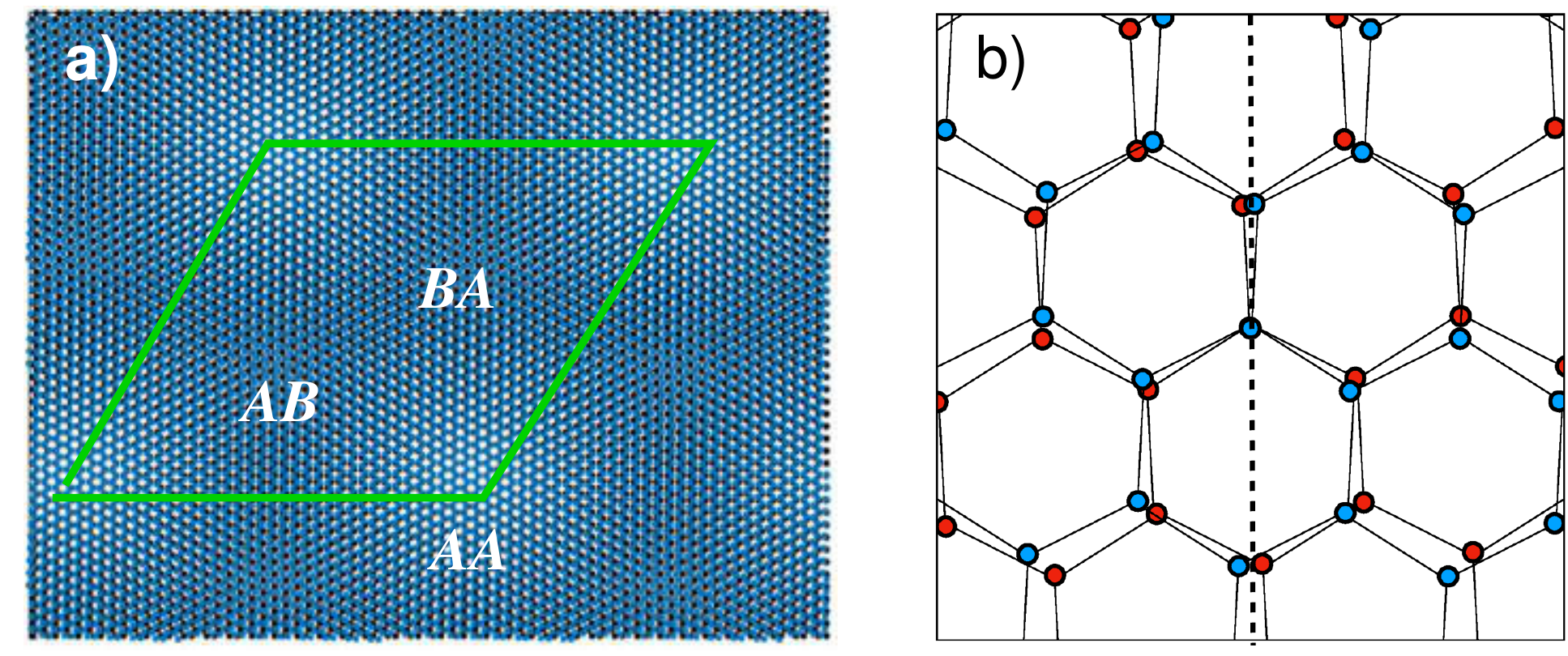}
\caption{{\small{}a) Moire pattern of twisted graphene bilayers for a twist
angle of  $\theta=1.8^{\circ}$. Each layer has two sublattices, $A$
and $B$. The pattern indicates regions of $AA$, $AB$, and $BA$
alignment. Four-fold degenerate states are observed around the $AA$
stacking regions. b) Twisted graphene bilayer rotated around $A$
sites ($AA$ region). At those points, the bilayer has $D_{3}$ symmetry,
comprised of a $C_{3}$ rotation around the $z$ axis and a $C_{2}^{\prime}$
rotation around the $y$ axis (dashed line) in between two layers. Red and blue
dots: top and bottom layer. }}
\end{figure}

After performing  calculations of the maximally localized Wannier orbitals
of the Moire superstructure, we establish the parameters of a \emph{minimal}
interacting tight-binding model that captures the Mott physics near
the magic angle. We show that even though the orbitals are well localized
in the Mott regime at quarter filling, surprisingly the direct \emph{exchange}
interaction between different sites is dominant and favors ferromagnetic
spin order at zero temperature. While charging effects \cite{Pizarro,Guinea},
which were not taken into account, may change our conclusions,
the scenario of zero temperature ferromagnetism in twisted graphene
bilayers seems in line with the reduced degeneracy of the Landau levels
measured with Shubnikov de Haas experiments near quarter filling
\cite{Cao1}. We discuss the experimental signatures of this state. 

\emph{Bloch Hamiltonian.$-$ }The free Hamiltonian for twisted graphene
bilayers can be constructed at the lattice level using a parametrization
for the hopping amplitudes between sites in the two different sheets,
\begin{equation}
\mathcal{H}=\left(\begin{array}{cc}
\mathcal{H}_{g}^{(1)} & \mathcal{H}_{\perp}\\
\mathcal{H}_{\perp}^{\dagger} & \mathcal{H}_{g}^{(2)}
\end{array}\right),\label{eq:Htwisted}
\end{equation}
where \emph{$\mathcal{H}_{g}$ }is the graphene Hamiltonian and \emph{$\mathcal{H}_{\perp}$
}is the interlayer hopping between the two sheets in real space. The
Moire pattern can be used to construct Bloch states that are periodic
in the superlattice vectors $\mathbf{T}_{i}$. For commensurate structures,
the Moire lattice vectors are parametrized by two integers $m$ and
$r$, and correspond to the twist angle $\cos\theta=1-r^{2}/2(3m^{2}+3mr+r^{2})$,
or equivalently $\theta\approx r/\sqrt{3}m$ for small angles. 

In a basis for Bloch states 
\begin{equation}
\Phi_{\mathbf{k},\sigma}\equiv(|\varphi_{\mathbf{k},A,\sigma}^{(1)}\rangle,|\varphi_{\mathbf{k},B,\sigma}^{(1)}\rangle,|\varphi_{\mathbf{k},A,\sigma}^{(2)}\rangle,|\varphi_{\mathbf{k},B,\sigma}^{(2)}\rangle)\label{basis}
\end{equation}
 defined in the two sublattices $A$ and $B$ of each of the two layers
$(1,2)$, the Bloch Hamiltonian of the twisted system
\begin{equation}
\mathcal{H}_{\mathbf{k}}(\mathbf{r},\mathbf{r}^{\prime})=\sum_{i}\mathcal{H}(\mathbf{r},\mathbf{r}^{\prime}+\mathbf{T}_{i})\text{e}^{i\mathbf{k}\cdot\mathbf{T}_{i}}\label{Bloch}
\end{equation}
satisfies $\mathcal{H}_{\mathbf{k}}(\mathbf{r},\mathbf{r}^{\prime}+\mathbf{T}_{i})=\mathcal{H}_{\mathbf{k}}(\mathbf{r},\mathbf{r}^{\prime}) e^{-i\mathbf{k}\cdot\mathbf{T}_i}$.
In that basis, 
\begin{equation}
[\mathcal{H}_{\mathbf{k}}]_{ab}=t_{\mathbf{k}}^{ab}(\mathbf{r},\mathbf{r}^{\prime})=\sum_{j}\text{e}^{i\mathbf{k}\cdot\mathbf{T}_{j}}t^{ab}(\mathbf{r},\mathbf{r}^{\prime}+\mathbf{T}_{j}),\label{eq:perp}
\end{equation}
are the matrix elements of (\ref{eq:Htwisted}), with $a,b$ indexes
running over the four components of basis (\ref{basis}).
The hopping amplitudes $t^{ab}(\mathbf{r},\mathbf{r}^{\prime})=\cos^{2}\theta_{z}V_{\sigma}(\mathbf{r}-\mathbf{r}^{\prime})+\sin^{2}\theta V_{\pi}(\mathbf{r}-\mathbf{r}^{\prime})$,
where $\cos\theta_{z}=d/\sqrt{d^{2}+(\mathbf{r}-\mathbf{r}^{\prime})^{2}}$
with $d$ the distance between the planes. $V_{\sigma}(\mathbf{r})$
and $V_{\pi}(\mathbf{r})$ are Slater-Koster functions \cite{V},
which decay exponentially and were parameterized following previous
ab initio works \cite{Lin,Tang}. 

Diagonalization of the Bloch Hamiltonian results in a
set of four-component Bloch eigenspinors $\hat{\psi}_{n,\mathbf{k}}(\mathbf{r})\equiv\langle\mathbf{r}|\hat{\psi}_{n,\mathbf{k}}\rangle$
that satisfy $\hat{\psi}_{n,\mathbf{k}}(\mathbf{r}+\mathbf{T})=\hat{\psi}_{n,\mathbf{k}}(\mathbf{r})\text{e}^{i\mathbf{k}\cdot\mathbf{T}}$
and correspond to the energy spectrum $\varepsilon_{n}(\mathbf{k})$. We calculate the bands for a small twist angle
of $\theta=1.0845^{\circ}$ ($m=30$, $r=1$) near the experimental
magic angle $\theta_{0}\sim1.1^{\circ}$. At that angle, the Bloch Hamiltonian is a $N_s \times N_s$ matrix with $N_s = 11164$ sites inside the Moire unit cell. The low energy bands ($n=1,\ldots,4$),  shown in Fig. 2b, are four-fold degenerate at the $K$
points (excluding the spin). They have a two-fold degeneracy at the
other two high symmetry points of the Brillouin zone, $\Gamma^{\prime}$
and $M^{\prime}$, where they open up a gap between particle and hole
branches. At the $\Gamma^{\prime}$ point, the Bloch states have $C_{3}$
and $C_{2y}^{\prime}$ symmetry, which involves at $\pi$ rotation
around the $y$ axis placed half-way between the two layers (shown
in Fig. 1b). We also find numerically that all Bloch eigenspinors 
satisfy the time reversal symmetry (TRS) relation $\mathcal{T}\hat{\psi}_{n,\mathbf{k}}(\mathbf{r})=\hat{\psi}_{n,-\mathbf{k}}^{*}(\mathbf{r})$,
with $\mathbf{k}$ measured from the center of the Moire Brillouin
zone at $\Gamma^{\prime}$.  The $K$ and $K^{\prime}$ points are
hence related by TRS, and must have opposite $\pi$ Berry phases.
This fact indicates that the Bloch states of the twisted structure
do not suffer from Wannier obstructions \cite{Po}, and hence could
be reconstructed through a proper basis of Wannier states. 

\begin{figure}[t]
\begin{centering}
\vspace{-0.3cm}\includegraphics[scale=0.44]{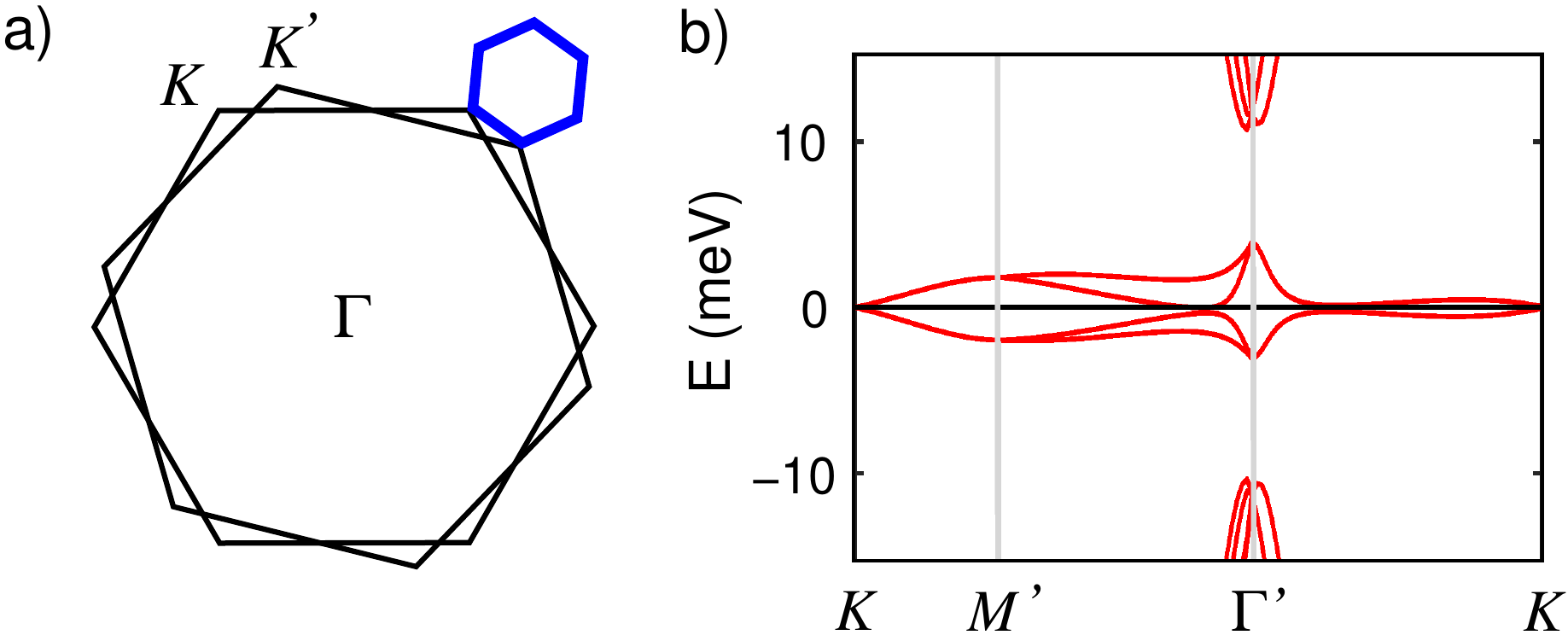}
\par\end{centering}
\caption{{\small{}a) Moire Brillouin zone of the twisted graphene system (blue
hexagon), containing the $K$ and $K^{\prime}$ points at the corners.
b) Flat bands in the Moire Brillouin zone for $\theta=1.0845^{\circ}$,
near the magic angle $\theta_{0}\sim1.1^{\circ}$. The $\Gamma^{\prime}$
point is at the center of the Moire Brillouin zone. $M^{\prime}$
is the mid point between $K$ and $K^{\prime}$ points. }}
\end{figure}

\emph{Wannier orbitals.$-$} From the Bloch states of the four low energy bands, one can extract
the Wannier wave functions in the Moire unit cell, 
\begin{equation}
|\mathbf{R}\nu\rangle=\frac{1}{N_s}\sum_{\mathbf{k}}\text{e}^{-i\mathbf{k}\cdot\mathbf{R}}U_{n\nu}(\mathbf{k})|\hat{\psi}_{n,\mathbf{k}}\rangle,\label{eq:R}
\end{equation}
where 
$\mathbf{R}$  is the center
of the Wannier orbitals and $U_{n\nu}(\mathbf{k})$  some $4 \times 4$ unitary
transformation. The four component Wannier spinors $\hat{W}_{\nu}(\mathbf{r}-\mathbf{R})\equiv\langle\mathbf{r}|\mathbf{R}\nu\rangle$
are not unique since adding a phase to the Bloch state $\text{e}^{-i\mathbf{k}\cdot\mathbf{r}} \hat{\psi}_{n\mathbf{k}}(\mathbf{r})$
corresponds to a new set of Wannier orbitals. We choose the set of
maximally localized Wannier orbitals in finding the unitary transformation
that minimizes their spread, $\Omega=\sum_{\nu}[\langle r^{2}\rangle_{\nu}-\langle r\rangle_{\nu}^{2}]$,
with $\langle X\rangle_{\nu}\equiv\langle\mathbf{R}\nu|X|\mathbf{R}\nu\rangle$.
The minimization was carried with the Wannier90 package \cite{Mostofi}.
The momentum space $\mathbf{k}$ mesh points are generated by the
reciprocal supercell lattice vectors with 300 $\times$ 300 grid points
using periodic boundary conditions, including all high symmetry points. 

Following the symmetry arguments outlined in Ref. \cite{Kang}, we
perform the minimization of the spread enforcing the $C_{3}$ and
$C_{2y}^{\prime}$ symmetry for the Bloch states around the $\Gamma^{\prime}$
points. Those two symmetries describe a $D_{3}$ point symmetry group,
which is a local symmetry of the lattice at $AA$ site regions when
the two graphene layers are rotated around a site \cite{Yuan}, as depicted in
Fig. 1b. In agreement with earlier results \cite{Kang,Koshino}, the
Wannier functions that satisfy those symmetries have three sharp peaks
centered around either the $AB$ or $BA$ sites, forming a honeycomb
superlattice with two-fold degenerate orbitals per site, as shown
in Fig. 3. 

On a given Moire unit cell, we label the Wannier orbitals by the four-component
spinors $\hat{W}_{\nu}=(w_{\nu,1},w_{\nu,2},w_{\nu,3},w_{\nu,4})^{T}$.
Among the four orbitals, $\hat{W}_{\nu}(r-\mathbf{R}_{j})$, two are
centered at $R_{j}\in AB$ sites and are eigenstates of the $C_{3}$
rotation operator, with eigenvalues $\epsilon=\text{e}^{2\pi i/3}$
and $\epsilon^{*}$. The other two are centered at $R_{j}\in BA$
sites and also have the same eigenvalues $\epsilon$ and $\epsilon^{*}$.
From now on, we will label the Wannier orbital spinors based on their
$C_{3}$ rotation eigenvalues, $C_{3}\hat{W}_{\alpha}(\mathbf{r}-\mathbf{R}_{j})=\text{e}^{\alpha2\pi i/3}\hat{W}_{\alpha}(\mathbf{r}-\mathbf{R}_{j})$,
with $\alpha=\pm$ and $R_{j}\in AB$ or $BA$. The two degenerate
orbitals centered at a given superlattice site $\mathbf{R}_{j}$ are
related by TRS, $\mathcal{T}\hat{W}_{\alpha}(\mathbf{r}-\mathbf{R}_{j})=\hat{W}_{-\alpha}(\mathbf{r}-\mathbf{R}_{j})$.
Orbitals in NN superlattice sites $\mathbf{R}_{i}$
and $\mathbf{R}_{j}$ are related by the $C_{2}^{\prime}$ rotation,
$C_{2}^{\prime}\hat{W}_{\alpha}(\mathbf{r}-\mathbf{R}_{i})=\hat{W}_{-\alpha}(\mathbf{r}-\mathbf{R}_{j})$.

\begin{figure}[t]
\begin{centering}
\vspace{-0.3cm}\includegraphics[scale=0.3]{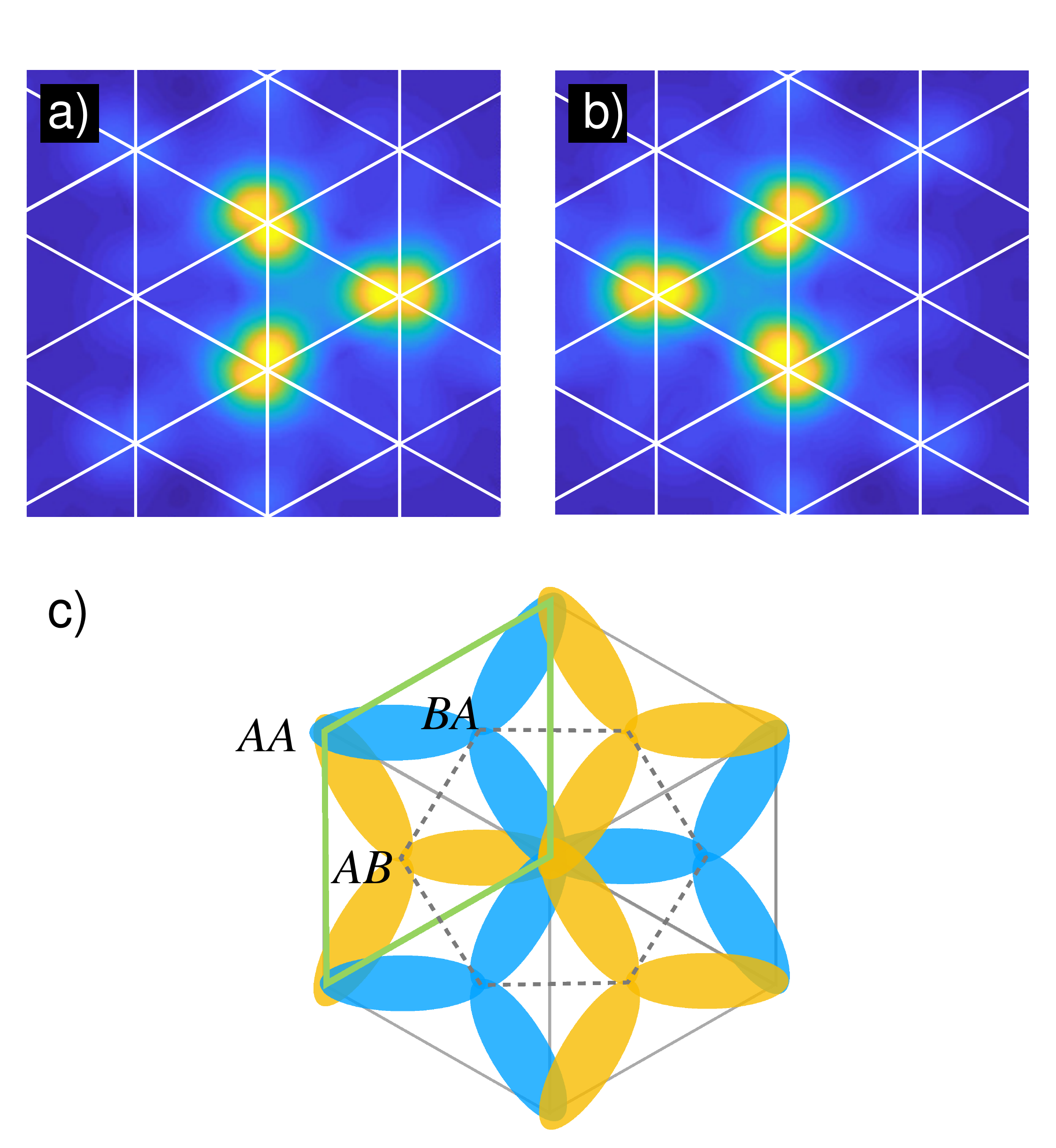}
\par\end{centering}
\caption{{\small{}Wannier wavefunction in the Moire superlattice. Amplitude
$|\hat{W}_{+}(\mathbf{r}-\mathbf{R}_{j})|=|\hat{W}_{-}(\mathbf{r}-\mathbf{R}_{j})|$
of the orbitals centered around (a) $j\in AB$ sites, and (b) $j\in BA$
sites, showing three sharp peaks. The orbitals $\hat{W}_{\alpha}$
have two-fold degeneracy per site, and are eigenstates of the $C_{3}$
rotation operator with eigenvalues $\epsilon$ ($\alpha=+$), and
$\epsilon^{*}$ ($\alpha=-$). c) Sketch of the orbitals in the Moire
unit cell (green line). Orange: AB centered Wannier orbitals. Blue:
$BA$ centered ones. The gray dotted line indicates the honeycomb
superlattice formed by the center of the orbitals. Their unusual three
peak structure indicates strong overlap between superlattice sites,
favoring ferromagnetic ordering at zero temperature. }}
\end{figure}

\emph{Tight binding Hamiltonian.$-$ }The effective lattice model
of this problem can be constructed by rewriting the Bloch Hamiltonian
(\ref{Bloch}) into a kinetic energy term of the form 
\begin{equation}
\mathcal{H}=\sum_{i,j}t_{\alpha\beta}(\mathbf{R}_{ij})d_{\alpha,\sigma}^{\dagger}(\mathbf{R}_{i})d_{\beta,\sigma}(\mathbf{R}_{j}),\label{eq:H0}
\end{equation}
where $\mathbf{R}_{i}$ indexes the sites of the  honeycomb superlattice,
$\mathbf{R}_{ij}\equiv\mathbf{R}_{i}-\mathbf{R}_{j}$ and the $d_{\alpha}(\mathbf{R})$
annihilates an electron with orbital of type $\alpha$ and spin $\sigma$
at a given superlattice site. The hopping matrix elements between
superlattice sites can be extracted from the matrix elements of Hamiltonian
(\ref{eq:Htwisted}) in a basis of maximally localized Wannier functions,
\begin{equation}
t_{\alpha\beta}(\mathbf{R}_{i})=\langle\mathbf{R}\alpha|\mathcal{H}|\mathbf{R}+\mathbf{R}_{i},\beta\rangle.\label{t}
\end{equation}
Due to the translational invariance of the superlattice, $t_{\alpha\beta}(\mathbf{R})=\langle0,\alpha|\mathcal{H}|\mathbf{R},\beta\rangle$.
For NN sites, we find that $|t_{\alpha\alpha}|(1)\approx0.384$ meV
whereas for $n$-th NN sites $t_{\alpha,-\alpha}(n)=0$. Hence, hopping between sites conserves the orbital
pseudospin quantum number $\alpha=\pm$. $|t_{\alpha\alpha}|(n)$ has a non-trivial dependence with the 
distance between sites (see table I), in qualitative agreement with the findings of Ref. \cite{Kang} for a significantly larger twist angle. 

\begin{table*}[t]
\begin{centering}
\begin{tabular}{rcrrrrrrr}
 &  &  &  &  &  &  &  & \tabularnewline
\hline 
\hline 
\vspace{-0.3cm} &  &  &  &  &  &  &  & \tabularnewline
(meV) &  & 0 & 1 & 2 & 3 & 4 & 5 & 6\tabularnewline
\vspace{-0.3cm} &  &  &  &  &  &  &  & \tabularnewline
\hline 
 &  & $\qquad\quad$ & $\qquad\qquad$ & $\qquad\qquad$ & $\qquad\qquad$ & $\qquad\qquad$ & $\qquad\qquad$ & $\qquad\qquad$\tabularnewline
|$t_{\alpha\alpha}|$ &  & 0 & 0.384 & 0.005 & 0.447 & 0.162 & 0.084 & 0.007\tabularnewline
\vspace{-0.3cm} &  &  &  &  &  &  &  & \tabularnewline
$V_{\alpha\beta}$ &  & 21.2 & 16.9 & 16.7 & 15.6 & 12.6 & 11.58 & 9.68\tabularnewline
\vspace{-0.3cm} &  &  &  &  &  &  &  & \tabularnewline
$J_{\alpha\alpha}$ &  & 0 & 5.09 & 1.11 & 0.52 & 0.25 & 0.16 & 0.09\tabularnewline
\vspace{-0.3cm} &  &  &  &  &  &  &  & \tabularnewline
$J_{\alpha,-\alpha}$ &  & 0 & $-$4.93 & 1.02 & $-$0.51 & $-$0.18 & 0.12 & 0.08\tabularnewline
\vspace{-0.2cm} &  &  &  &  &  &  &  & \tabularnewline
\hline 
\hline 
 &  &  &  &  &  &  &  & \tabularnewline
\end{tabular}
\par\end{centering}
\caption{{\small{}Electronic hopping amplitude $|t_{\alpha\alpha}|$, direct
Coulomb interaction $V_{\alpha\beta}$ and exchange interaction $J_{\alpha\beta}$
for various nearest neighbor sites: on-site ($0$), and $n$-th nearest
neighbors ($n$), with $n=1-6$. Energies in meV calculated for $\kappa=5$.
$n=1,\:3$ and $4$ correspond to sites in opposite sublattices. $J_{\alpha\alpha}\approx\pm J_{\alpha,-\alpha}$,
with $+$($-$) for sites in the same (opposite) sublattice. }}
\end{table*}

The Coulomb interactions between lattice sites can be written as $\mathcal{H}_{C}=\frac{1}{2}\int\text{d}\text{\textbf{r}}\text{d}\mathbf{r}^{\prime}\rho(\mathbf{r})\frac{e^{2}}{\kappa|\mathbf{r}-\mathbf{r}^{\prime}|}\rho(\mathbf{r}^{\prime})$,
where $\rho(\mathbf{r})$ is the electron density and $\kappa\approx5$
the dielectric constant of twisted bilayers encapsulated in boron
nitride. We can rewrite this term in terms of $d_{\alpha,\sigma}$
operators by expressing the density $\rho(\mathbf{r})=\sum_{\sigma}\hat{\Psi}_{\sigma}^{\dagger}(\mathbf{r})\hat{\Psi}_{\alpha}(\mathbf{r})$
in terms of field operators $\hat{\Psi}_{\sigma}(\mathbf{r})=\sum_{\alpha,j}\hat{W}_{\alpha}(\mathbf{r}-\mathbf{R}_{j})d_{\alpha,\sigma}.$
The resulting Coulomb Hamiltonian has a direct term and also an exchange
part, $\mathcal{H}_{C}=\mathcal{H}_{d}+\mathcal{H}_{e}$. The first
term, 
\begin{equation}
\mathcal{H}_{d}=\sum_{i,j}V_{\alpha\beta}(\mathbf{R}_{ij})n_{\alpha}(\mathbf{R}_{i})n_{\beta}(\mathbf{R}_{j}),\label{eq:}
\end{equation}
with $n_{\alpha}(\mathbf{R})=\sum_{\sigma}d_{\alpha,\sigma}^{\dagger}(\mathbf{R})d_{\alpha,\sigma}(\mathbf{R})$
the density operator and repeated $\alpha,\,\beta$ indexes to be
summed. The Coulomb coupling is cast as an overlap integral of Wannier
orbital spinors, $V_{\alpha\beta}(\mathbf{R}_{ij})=\frac{1}{2}\int\text{d}\mathbf{r}\text{d}\mathbf{r}^{\prime}|\hat{W}_{\alpha}(\mathbf{r}_{i})|^{2}\frac{e^{2}}{\kappa|\mathbf{r}-\mathbf{r}^{\prime}|}|\hat{W}_{\beta}(\mathbf{r}_{j}^{\prime})|^{2},$
with $\mathbf{R}_{ij}=\mathbf{R}_{i}-\mathbf{R}_{j}$ and $\mathbf{r}_{j}\equiv\mathbf{r}-\mathbf{R}_{j}$.
The exchange part is 
\begin{equation}
\mathcal{H}_{e}=\sum_{i\neq j}J_{\alpha\alpha^\prime, \beta\beta^\prime}(\mathbf{R}_{ij})\,d_{\alpha,\sigma}^{\dagger}(\mathbf{R}_{i})d_{\beta,\sigma^{\prime}}^{\dagger}(\mathbf{R}_{j})d_{\beta^\prime,\sigma^{\prime}}(\mathbf{R}_{i})d_{\alpha^\prime,\sigma}(\mathbf{R}_{j}),\label{He}
\end{equation}
where
\begin{align}
J_{\alpha\alpha^\prime,\beta \beta^\prime}(\mathbf{R}_{ij}) & =\frac{1}{2}\int\text{d}\mathbf{r}\text{d}\mathbf{r}^{\prime}\hat{W}_{\alpha}^{\dagger}(\mathbf{r}_{i})\cdot\hat{W}_{\alpha^\prime}(\mathbf{r}_{j})\frac{e^{2}}{\kappa|\mathbf{r}-\mathbf{r}^{\prime}|}\nonumber \\
 & \qquad\qquad\times\hat{W}_{\beta}^{\dagger}(\mathbf{r}_{j}^{\prime})\cdot\hat{W}_{\beta^\prime}(\mathbf{r}_{i}^{\prime})\label{J}
\end{align}
is the exchange coupling between lattice sites. In general, we find that 
the combinations $J_{\alpha\beta,\beta\alpha}(\mathbf{R}_{ij})=  J_{\alpha\beta,\alpha\beta}(\mathbf{R}_{ij})=0$ for $\alpha\neq\beta$, within the numerical precision. 
That includes the on site exchange  (Hund's coupling), which is zero due to the orthogonality
between same site Wannier spinors \cite{Uchoa1, Koshino}. From now on, we define the {\it only} non-zero combination
$J_{\alpha\alpha, \beta\beta}\equiv J_{\alpha\beta}$.

The numerical values of the hopping energy, Coulomb interaction and
the exchange interaction for $n$-th NNs, is shown in table I, which
is the first main result of the paper. We find the on-site Hubbard $U_{\alpha\beta}\equiv V_{\alpha\beta}(0)=21.2$
meV, which is much larger than the first NN hopping $t(1)$, and hence
the ratio $U/t(1)\sim55$ falls comfortably in the realm of the Mott
regime. 

The exchange interaction for first NN sites $(n=1)$ is $|J_{\alpha\beta}(1)|\approx5$
meV. In general, the diagonal terms $J_{\alpha\alpha}(n)>0$ are positive
definite, whereas the off diagonal ones can be either positive or
negative, $J_{\alpha,-\alpha}(n)\approx\pm J_{\alpha\alpha}(n)$,
with $+$ ($-$) for $i,\,j$ sites in the same (opposite) sublattice, as shown in table I.
For sites in the same sublattice, the fact that $J_{\alpha\beta}(n)\approx J_{\alpha\alpha}(n)>0$
is the same for all four combinations of $\alpha,\beta=\pm$ indexes
hints at an emergent SU(4) symmetry between spin and orbital degrees
of freedom at quarter filling. For sites in opposite sublattices,
the exchange interaction has SU(2) symmetry in the spin. It has also
both ferro $(J_{\alpha\alpha}>0$) and antiferromagnetic $(J_{\alpha,-\alpha}<0)$
correlations in the orbital sector, depending on the orientation of
the pseudospins. 

Since Hund's coupling is zero, at quarter filling
the lower flat bands are in the unitary limit \cite{Coqblin}, with
each Moire superlattice site \textbf{$\mathbf{R}_{j}$} being singly
occupied and having a well defined spin $\sigma$ and orbital quantum
number $\alpha=\pm$. Mapping the exchange term in terms of spin $\mathbf{S}_{i}=\frac{1}{2}d_{\alpha,\sigma}^{\dagger}(\mathbf{R}_{i})\vec{\sigma}_{\sigma\sigma^{\prime}}d_{\alpha,\sigma^{\prime}}(\mathbf{R}_{i})$
and pseudospin $\boldsymbol{\tau}_{i}=\frac{1}{2}d_{\alpha,\sigma}^{\dagger}(\mathbf{R}_{i})\vec{\sigma}_{\alpha\beta}d_{\beta,\sigma}(\mathbf{R}_{i})$
operators, the result is the ferromagnetic exchange interaction announced
in Eq. (1), with $J_{ij}\equiv J_{\alpha\alpha}(n)>0$ \cite{SM1}.
This Hamiltonian favors ferromagnetic alignment of the spins at zero
temperature ($T=0$). In the orbital sector different  states
are possible, 
 including canted magnetism with ferromagnetic 
order in the pseudospin $\tau^{z}$ component, accompanied by 
staggered (antiferromagnetic) order in the transverse, $\boldsymbol{\tau}^{\perp}$
direction.

The superexchange interaction follows from second order perturbation
theory in the hopping energy \cite{kk,Vanderbos} and has the same
form as the exchange term in Eq. (1) for $\eta_{ij}=-1$ \cite{Uchoa1}.
The superexchange term has SU(4) symmetry and favors antiferromagnetic
alignment between nearest neighbor sites due to Pauli principle. It's
coupling $J\to-t^{2}/U\approx-0.01$ meV is very small compared to
the exchange one, and can be safely igonored. 

\emph{Ferromagnetic Mott state.$-$} Mott-Hubbard insulators have
strongly localized states and are known to be overwhelmingly antiferromagnetic
due to strong superexchange interactions ($t^{2}/U\gg J$) \cite{khomskii,anderson,anderson-2}.
Ferromagnetism occurs mostly either in metallic systems or in metallic
bands hybridized with localized moments via the Anderson impurity
mechanism \cite{anderson,anderson-2,moriya}. Within the Hubbard model
framework, 
the only credible mechanism for spin ferromagnetism
exists for multi-orbital systems in the context of the Kugel-Khomskii
model \cite{khomskii,kk}, where superexchange can become effectively
ferromagnetic in the presence of staggered orbital ordering. 

We conjecture that the flat bands in twisted graphene bilayers are
in a way intermediate between ferromagnetic bad metals and antiferromagnetic
Mott-Hubbard insulators. Due to the exotic shape of the Wannier orbitals,
the hierarchy between hopping, direct exchange and the local Hubbard
interaction, $t\ll J\ll U,$ leads to an anomalously small superexchange.
In the charge sector the Mott gap is also anomalously small,  $\Delta\sim0.3\text{ meV}\ll W\ll U$,
 where $W\sim5$ meV  is the bandwidth,  and the system undergoes an insulator-metal transition
at $T\approx4$K \cite{Cao1}. 

In spite of the fact that $U/t$ is large, the strong overlap between
the orbitals found in the non-interacting theory suggests that the
system is potentially close to an insulator-metal transition \cite{imada}
due to a charge fluctuation mechanism which presently is not well-understood
\cite{Pizarro,Guinea}. Nevertheless, the effective spin model we
propose in this work should not depend on the details of this mechanism,
as long as the system remains quarter filled and does not undergo
a charge-ordering transition (potentially accompanied by dimerization)
due to Coulomb interactions. In carbon lattices, which are notoriously
stiff \cite{Lee}, charge density wave instabilities are hindered
by the high elastic energy cost for the system to deform the lattice
and restore charge neutrality. 

\emph{Experimental signatures.$-$} Since the honeycomb superlattice
is not frustrated, it will exhibit ferromagnetic spin order at $T=0$
in the universality class of the ferromagnetic (spin $S$) Heisenberg
model. It is well known that the magnetization $M$, correlation length
$\xi$ and the spin susceptibility $\chi$ exhibit peculiar features
in two dimensions, since for any $T\neq0$ the system is disordered, with zero Curie temperature. 
The model has been extensively studied
both in zero and finite external magnetic field $H$ on various lattices \cite{kopietz,antsygina,junger,note2}. 
At finite field $H\neq0$, $M(H)$ is finite and strongly temperature
dependent. In the regime $T/J\ll1$, which can take place for $T\approx2$K
(where $T/J\approx1/25$), a weak magnetic field of $H\approx0.2$T
(i.e. $H/J\approx1/250)$ already provides nearly maximum magnetization
\cite{antsygina,junger}. The susceptibility $\chi(H)$ is zero for
$T=0$ and $H\neq0$ and exhibits a characteristic finite-temperature
peak at $T=T_{\chi}$ which scales in a well-defined way with external
field. 

It has been established experimentally that doping away from the Mott
insulating phase leads to metallic (and even superconducting) behavior \cite{Cao1,Cao2}. 
Therefore the structure of the ground state and excitation spectrum of this unconventional metallic state is of great 
experimental and theoretical interest. A profound new feature has emerged at finite magnetic field, which
persists both in weak  (Shubnikov-de Haas oscillations) and strong field limits  (Quantum Hall effect), for hole doping \cite{Cao1,Cao2}.
 Those measurements suggest  a small Fermi surface that develops from doping the correlated insulating phase, accompanied by a possible symmetry
 breaking of yet unknown origin. The resulting state has a fermionic  degeneracy  of 2, indicating  a reduction of the original 
 four-fold band degeneracy by a factor of 2.   
 
 This behavior is consistent with the system being in the proximity to a ferromagnetic Mott state, in which the spins align when nudged by an infinitesimally weak field. At the same time, any long-range order in the orbital sector is expected to be much more fragile and disappear  quickly  due to charge disorder and motion of holes in the metallic state. 
 Therefore we conjecture that in the weak field limit,  the ground state emerging
  from doping the ferromagnetic insulator would be a ferromagnetic, spin-polarized, strongly-correlated metal,
with the orbital pseudospin symmetry preserved. 

\emph{Conclusions.$-$ }We  have  derived the effective spin model that describes
the Mott phase of twisted graphene bilayers at the magic angle. After calculating
the maximally localized Wannier wavefunctions from the lattice, we
propose that the system forms a novel \emph{ferromagnetic} Mott state
at quarter filling, with clear experimental signatures. 

\emph{Acknowledgements.}$-$ BU Acknowledges P. Jarillo-Harillo, T.
Senthil, and K. Beach for discussions. K.S. and B. U. acknowledge
NSF CAREER grant No. DMR-1352604 for support.

{\it Note added}.$-$  After the submission of this work, we became aware of Ref.~\cite{Oskar}, which also found a ferromagnetic ground state using different arguments. We thank O. Vafek for pointing it out.

\newpage


\section*{Supplementary material}

# Supplementary Materials of "Ferromagnetic Mott state in twisted graphene bilayers at the magic angle"

Kangjun Seo, Valeri N. Kotov and Bruno Uchoa

*Department of Physics and astronomy, University of Oklahoma, Norman, OK 73019, USA and*
*Department of Physics, University of Vermont, Burlington, VT 05405, USA*

## MOIRÉ SUPERLATTICE

Each graphene layer consists of two sublattices ($A_1$, $B_1$ in layer 1 and $A_2$, $B_2$ in layer 2). The positions of the carbon atoms in each sublattice in layer 1 (bottom layer) are

$$\mathbf{r}_{B_1} = n\,\mathbf{a}_1 + m\,\mathbf{a}_2, \quad \mathbf{r}_{A_1} = \mathbf{r}_{B_1} + \boldsymbol{\delta}_1, \tag{1}$$

where $n$ and $m$ are integers, $\boldsymbol{\delta}_1 = 2/3(\mathbf{a}_1 + \mathbf{a}_2)$, and $\mathbf{a}_j$ are basis vectors of the unit cell,

$$\mathbf{a}_1 = a\,(1,0), \quad \mathbf{a}_2 = a\,(1,\sqrt{3})/2 \tag{2}$$

with the lattice parameter $a = 2.46$ Å. When the layers are not twisted ($\theta = 0$), the horizontal position of site $B_2$ is the same as of site $B_1$:

$$\mathbf{r}_{B_2} = n\,\mathbf{a}_1 + m\,\mathbf{a}_2, \quad \mathbf{r}_{A_2} = \mathbf{r}_{B_2} + \boldsymbol{\delta}_1, \tag{3}$$

After the rotation, the lattice vectors of layer $l$ are given by

$$\mathbf{a}_i(l) = R(\pm\theta/2)\,\mathbf{a}_i(l), \quad R(\theta) = \begin{pmatrix} \cos\theta & \sin\theta \\ -\sin\theta & \cos\theta \end{pmatrix} \tag{4}$$

with $R(\theta)$ being the rotation matrix around $B_1$ and $B_2$. Then the reciprocal lattice vectors becomes

$$\mathbf{b}_i(l) = R(\pm\theta/2)\,\mathbf{b}_i, \tag{5}$$

where $\mathbf{b}_1 = \frac{2\pi}{a}\,(1,-1/\sqrt{3})$ and $\mathbf{b}_2 = \frac{2\pi}{a}\,(0, 2/\sqrt{3})$.

The lattice is commensurate if [1, 3–5]

$$\cos\theta = \frac{3m_0^2 + 3m_0 r + r^2/2}{3m_0^2 + 3m_0 r + r^2}, \tag{6}$$

where $m_0$ and $r$ are coprime positive integers. The superlattice vectors $\mathbf{T}_{1,2}$ can be expressed in terms of $m_0$, $r$, and $\mathbf{a}_{1,2}$. These expressions are different when $r$ is either divisible or nondivisible by 3. The number of sites in each supercell is

$$N(m_0, r) = \begin{cases} 4(3m_0^2 + 3m_0 r + r^2), & r \neq 3n, \\ 4(3m_0^2 + 3m_0 r + r^2)/3, & r = 3n. \end{cases} \tag{7}$$

The linear size of the superlattice cell is

$$|\mathbf{T}_1| = |\mathbf{T}_2| = a\sqrt{N}/2. \tag{8}$$

The Moiré period $L$ is related to the twist angle,

$$L = \frac{a}{2\sin(\theta/2)}. \tag{9}$$

In our numerical calculations, we use $m_0 = 30$ and $r = 1$ which correspond to $\theta = 1.0849°$, $L = 129.9196$ Å, and $N = 11164$. The superlattice vectors and the corresponding reciprocal superlattice vectors are

$$\mathbf{T}_1 = L\,(0,-1), \quad \mathbf{T}_2 = L\,(\sqrt{3}/2, -1/2). \tag{10}$$

$$\mathbf{G}_1 = \frac{2\pi}{L}(-1/2, -\sqrt{3}/2), \quad \mathbf{G}_2 = \frac{2\pi}{L}(1, 0). \tag{11}$$

The two nonequivalent Dirac points are

$$\mathbf{K}_1 = \frac{\mathbf{G}_1 + 2\mathbf{G}_2}{3}, \quad \mathbf{K}_2 = \frac{2\mathbf{G}_1 + \mathbf{G}_2}{3}. \tag{12}$$

The existence of a commensuration depends only on the relative rotation of the lattice vectors of each layer, and not on the structure of the unit cells of each layer. These amount to different choices of initial basis vectors within each cell. [2]

### MICROSCOPIC HAMILTONIAN OF GRAPHENE BILAYER

It is convenient to enumerate the sites in the sublattice in each layer using two integer valued vectors $\mathbf{j} = (i, j)$ and $\mathbf{x} = (n, m)$, where $\mathbf{j}$ labels the positions of the supercell and $\mathbf{x}$ enumerates the sites in the supercell. Then, the Hamiltonian has the form

$$\mathcal{H} = \sum_{\langle \mathbf{ix}, \mathbf{jy}\rangle, l=1,2,\sigma} \left[ t(\mathbf{T_i} + \mathbf{r}_{\mathbf{x}_{a_l}}, \mathbf{T_j} + \mathbf{r}_{\mathbf{y}_{a'_l}}) \, c_\sigma^\dagger(\mathbf{i}, \mathbf{x}_{a_l}) \, c_\sigma(\mathbf{j}, \mathbf{y}_{a'_l}) + \text{h.c.} \right]$$
$$+ \sum_{\langle \mathbf{ix}, \mathbf{jy}\rangle, a, a', \sigma} \left[ t(\mathbf{T_i} + \mathbf{r}_{\mathbf{x}_{a_1}}, \mathbf{T_j} + \mathbf{r}_{\mathbf{y}_{a'_2}}) \, c_\sigma^\dagger(\mathbf{i}, \mathbf{x}_{a_1}) \, c_\sigma(\mathbf{j}, \mathbf{y}_{a'_2}) + \text{h.c.} \right], \tag{13}$$

where $\mathbf{T_j} = i\mathbf{T}_1 + j\mathbf{T}_2$, and $c_\sigma^\dagger(\mathbf{i}, \mathbf{x}_{a_l})$ is the creation operator of an electron with the spin $\sigma$ on the sublattice $a_l$ ($= A_l, B_l$) in the supercell $\mathbf{i}$ in the position $\mathbf{x}$ on layer $l$.

Note that the Hamiltonian is invariant with respect to translation by $\mathbf{T}_{1,2}$, so we can perform the Fourier transform

$$\varphi_{\mathbf{k},a,\sigma}^{(l)}(\mathbf{x}) = \frac{1}{\sqrt{N_s}} \sum_{\mathbf{j}} e^{i\mathbf{k}\cdot\mathbf{T_j}} c_\sigma(\mathbf{i}, \mathbf{x}_{a_l}), \tag{14}$$

where $N_s$ is the number of the supercells. Note that

$$t(\mathbf{T_i} + \mathbf{r_x}, \mathbf{T_j} + \mathbf{r_y}) = t(\mathbf{T_i} - \mathbf{T_j} + \mathbf{r_x} - \mathbf{r_y}), \tag{15}$$

so that we have the Bloch Hamiltonian

$$\mathcal{H}_\mathbf{k}(\mathbf{x}, \mathbf{y}) = \sum_{\mathbf{i}} \mathcal{H}(\mathbf{x}, \mathbf{y} + \mathbf{T_i}) \, e^{i\mathbf{k}\cdot\mathbf{T_i}}. \tag{16}$$

Introducing the $N$-component operator

$$\Phi_{\mathbf{k}\sigma}(\mathbf{x}) = \left[ \varphi_{\mathbf{k},A,\sigma}^{(1)}, \varphi_{\mathbf{k},B,\sigma}^{(1)}, \varphi_{\mathbf{k},A,\sigma}^{(2)}, \varphi_{\mathbf{k},B,\sigma}^{(2)} \right]^T, \tag{17}$$

we have the Hamiltonian in the form $H = 1/N_s \sum_{\mathbf{k}\sigma} \sum_{\mathbf{x},\mathbf{y}} \Phi_{\mathbf{k}\sigma}^\dagger(\mathbf{x}) H_\mathbf{k}(\mathbf{x}, \mathbf{y}) \Phi_{\mathbf{k}\sigma}(\mathbf{y})$, with the matrix elements indicated in Eq. (5) of the main text.

The energy spectrum $\varepsilon_{n\mathbf{k}}$ and the corresponding Bloch state $|\hat\psi_{n,\mathbf{k},\sigma}\rangle$ with $n$ band indices can be obtained by diagonalizing the above Hamiltonian matrix:

$$\sum_{\mathbf{y}} \mathcal{H}_\mathbf{k}(\mathbf{x}, \mathbf{y}) \langle \mathbf{y} | \hat\psi_{n,\mathbf{k},\sigma}\rangle = \varepsilon_{n\mathbf{k}} \langle \mathbf{x} | \hat\psi_{n,\mathbf{k},\sigma}\rangle. \tag{18}$$

Note that the Bloch wavefunctions $\hat\psi_{n,\mathbf{k},\sigma}(\mathbf{r_x}) = \langle \mathbf{x} | \hat\psi_{n,\mathbf{k},\sigma}\rangle$ satisfy

$$\hat\psi_{n,\mathbf{k},\sigma}(\mathbf{r_x} + \mathbf{T_i}) = e^{i\mathbf{k}\cdot\mathbf{T_i}} \hat\psi_{n,\mathbf{k},\sigma}(\mathbf{r_x}), \tag{19}$$

and the orthogonality

$$\langle \hat\psi_{n,\mathbf{k},\sigma} | \hat\psi_{n',\mathbf{p},\sigma}\rangle = \sum_{\mathbf{ix}} \hat\psi_{n,\mathbf{k},\sigma}^\dagger(\mathbf{r_x} + \mathbf{T_i}) \hat\psi_{n',\mathbf{k},\sigma}(\mathbf{r_x} + \mathbf{T_i}) = N_s \, \delta_{\mathbf{k},\mathbf{p}} \delta_{n,n'}. \tag{20}$$

Since the periodicity of the Bloch wavefunctions, we will use $\mathbf{r} \equiv \mathbf{r_x} + \mathbf{T_j}$ for the atomic positions in the entire lattice. In our calculation, the momentum mesh points $(k_1, k_2)$ is generated by the reciprocal superlattice vectors $\mathbf{G}_1$ and $\mathbf{G}_2$ by $\mathbf{k} = k_1 \mathbf{G}_1 + k_2 \mathbf{G}_2$ using periodic boundary condition, and we used $300 \times 300$ grid points including $\Gamma$, $K_1$, and $K_2$ in the mini Brillouin zone.

# WANNIER FUNCTIONS

The Wannier function centered at $\mathbf{R}$ associated with an energy band $n$ is

$$|\mathbf{R}n\rangle = \frac{1}{N} \sum_{\mathbf{k}} e^{-i\mathbf{k}\cdot\mathbf{R}} |\hat{\psi}_{n\mathbf{k}}\rangle. \tag{21}$$

However, the Wannier functions are not unique, so given one set of of Bloch orbitals, another equally good set is obtained from

$$|\hat{u}_{\mu\mathbf{k}}\rangle \to \sum_n U_{n\mu}(\mathbf{k}) |\hat{u}_{n\mathbf{k}}\rangle, \tag{22}$$

where $|u_{n\mathbf{k}}\rangle = e^{-i\mathbf{k}\cdot\mathbf{r}} |\psi_{n\mathbf{k}}\rangle$ is the periodic part of the Bloch function, and $U_{n\mu}(\mathbf{k})$ is a unitary matrix that mixes the bands at $\mathbf{k}$:

$$\sum_n U^*_{n\mu}(\mathbf{k}) \, U_{n\mu}(\mathbf{k}) = \delta_{\mu\nu}. \tag{23}$$

The transformation does not preserve the individual Wannier centers, but does preserve the the sum of the the Wannier centers, modulo a lattice vector $\mathbf{T}$. Therefore, the general Wannier function can be written as

$$|\mathbf{R}\mu\rangle = \frac{1}{N} \sum_{\mathbf{k},n} e^{-i\mathbf{k}\cdot\mathbf{R}} U_{n\mu}(\mathbf{k}) |\hat{\psi}_{n\mathbf{k}}\rangle, \tag{24}$$

or, we can define $W_\mu(\mathbf{r} - \mathbf{R})$ as a localized function at $\mathbf{R}$:

$$\hat{W}_\mu(\mathbf{r} - \mathbf{R}) = \langle \mathbf{r} | \mathbf{R}\mu \rangle = \frac{1}{N} \sum_{\mathbf{k},n} e^{-i\mathbf{k}\cdot\mathbf{R}} U_{n\mu}(\mathbf{k}) \, \hat{\psi}_{n\mathbf{k}}(\mathbf{r}). \tag{25}$$

Note that the Wannier function translated by $\mathbf{T_i}$ is orthogonal: Since

$$\begin{aligned} \hat{W}_\mu(\mathbf{r} - \mathbf{R} + \mathbf{T_i}) &= \langle \mathbf{r} + \mathbf{T_i} | \mathbf{R}\mu \rangle = \langle \mathbf{r} | \mathbf{R} - \mathbf{T_i}, \mu \rangle \\ &= \frac{1}{N} \sum_{\mathbf{k},n} e^{-i\mathbf{k}\cdot(\mathbf{R}-\mathbf{T_i})} U_{n\mu}(\mathbf{k}) \, \hat{\psi}_{n\mathbf{k}}(\mathbf{r}), \end{aligned} \tag{26}$$

we have

$$\begin{aligned} \langle \mathbf{R}\mu | \mathbf{R} - \mathbf{T_i}, \nu \rangle &= \frac{1}{N^2} \sum_{\mathbf{k},\mathbf{p},n,n'} U^*_{n\mu}(\mathbf{k}) U_{n'\nu}(\mathbf{p}) \, e^{i\mathbf{p}\cdot\mathbf{T_i}} \langle \hat{\psi}_{n\mathbf{k}} | \hat{\psi}_{n',\mathbf{p}} \rangle \\ &= \frac{1}{N} \sum_{\mathbf{k},n} U^*_{n\mu}(\mathbf{k}) U_{n\nu}(\mathbf{k}) \, e^{i\mathbf{k}\cdot\mathbf{T_i}} = \frac{1}{N} \sum_{\mathbf{k}} \delta_{\mu\nu} \, e^{i\mathbf{k}\cdot\mathbf{T_i}} \\ &= \delta_{\mu\nu} \, \delta_{\mathbf{T_i},\mathbf{0}}. \end{aligned}$$

As a measure of the delocalization or spread of the Wannier functions, $\Omega[U(\mathbf{k})] = \sum_\mu \left[ \langle r^2 \rangle_\mu - \langle r_\mu \rangle^2 \right]$ with $\langle r_\mu \rangle = \langle \mathbf{0}\mu | r | \mathbf{0}\mu \rangle$ and $\langle r^2 \rangle_\mu = \langle \mathbf{0}\mu | r^2 | \mathbf{0}\mu \rangle$, the maximally localized Wannier functions can be obtained by minimizing $\Omega$ with respect to $U_{n\mu}(\mathbf{k})$ [8].

# SITE-SYMMETRY GROUP AND SYMMETRY-ADAPTED WANNIER FUNCTIONS

Specifying a set of sites $\{\mathbf{R}_1, \mathbf{R}_2, \cdots\}$, where the Wannier functions will be centered, the site-symmetry group of a given $\mathbf{R}_1$, denoted by $G_{\text{site}}$, is a subgroup of the full symmetry group of the lattice, $G$ whose elements leave $\mathbf{R}_1$ unchanged, so $g_{\text{site}} = (R_s | \mathbf{T}_s) \in G_{\text{site}}$ satisfy

$$g_{\text{site}} \mathbf{R}_1 = (R_s | \mathbf{T}_s) \mathbf{R}_1 = R_s \mathbf{R}_1 + \mathbf{T}_s = \mathbf{R}_1, \tag{27}$$

where $R_s$, $\mathbf{T}_s$ are the rotation and the translation part of the symmetry operation. The full symmetry group $G$ can be decomposed into left cosets of the subgroup $G_{\text{site}}$ as

$$G = \sum_{j,n} g_{jn} G_{\text{site}}, \quad g_{jn} = (R_j | \mathbf{T}_j + \mathbf{T}_n). \tag{28}$$

Here, $g_{j0}$ ($\mathbf{T}_0 = \mathbf{0}$) is one of the symmetry operations that maps $\mathbf{R}_1$ to its symmetry-equivalent point $\mathbf{R}_j$ as

$$\mathbf{R}_j = g_{j0} \mathbf{R}_1 = R_j \mathbf{R}_1 + \mathbf{T}_j. \tag{29}$$

Since $j = 1$ corresponds to the original point $\mathbf{R}_1$, $g_{10} = (E|\mathbf{0})$, where $E$ denotes the identity operation.

From the site-symmetry group for a given $\mathbf{R}_1$, the site-symmetry adapted Wannier functions centered at $\mathbf{R}_1$ are defined as the basis functions of the irreducible representations of the site-symmetry group $G_{\text{site}}$. These Wannier functions are represented as

$$\hat{W}_\alpha(\mathbf{r}_j) \equiv \hat{W}_\alpha(\mathbf{r} - \mathbf{R}_j), \tag{30}$$

where $\alpha = 1, 2, \cdots, n_\alpha$ denote the component of the basis functions with $n_\alpha$ the dimension of the irreducible representation. The Wannier functions transform as

$$\hat{g}_{\text{site}} \hat{W}_\alpha(\mathbf{r}) = \hat{W}_\alpha(g_{\text{site}}^{-1} \mathbf{r} - \mathbf{R}_1) = \sum_{\alpha'=1}^{n_\alpha} d_{\alpha'\alpha}(R_s) \hat{W}_{\alpha'}(\mathbf{r}_1) \tag{31}$$

From $\hat{W}_{\alpha 1}(\mathbf{r})$, we can generate Wannier functions centered at $\mathbf{R}_j$ as

$$\hat{W}_\alpha(\mathbf{r}_j) = \hat{g}_{j0} \hat{W}_\alpha(\mathbf{r}_1). \tag{32}$$

In our model, $\hat{g}_{\text{site}} = \hat{C}_3$ and $\mathbf{R}_1$ is a site $BA$, leading to $\hat{W}_1(\mathbf{r}_1) = \hat{W}_+(\mathbf{r}_1) = \hat{W}_+(\mathbf{r} - \mathbf{R}_1)$ and $\hat{W}_2(\mathbf{r}_1) = W_-(\mathbf{r}_1) = W_-(\mathbf{r} - \mathbf{R}_1)$ with

$$d_{\alpha'\alpha} = \begin{pmatrix} \epsilon & 0 \\ 0 & \epsilon^* \end{pmatrix}, \quad \alpha', \alpha = 1, 2, \quad \epsilon = \exp(2\pi i/3). \tag{33}$$

For $\mathbf{R}_2$ a site $AB$, we have $\hat{W}_1(\mathbf{r}_2) = \hat{W}_-(\mathbf{r}_2) = \hat{W}_-(\mathbf{r} - \mathbf{R}_2)$ and $\hat{W}_2(\mathbf{r}_2) = \hat{W}_+(\mathbf{r}_2) = \hat{W}_+(\mathbf{r} - \mathbf{R}_2)$ with $R_2 = C'_{2y}$.

From these symmetry-adapted Wannier functions $W_{\alpha j}(\mathbf{r})$, one can construct the Bloch functions $\Psi_{\mathbf{k},\alpha j}(\mathbf{r})$ as

$$\hat{\psi}_{\mathbf{k},\alpha}(\mathbf{r}_j) = \sum_n e^{i\mathbf{k}\cdot\mathbf{T}_n} \hat{W}_\alpha(\mathbf{r}_j - \mathbf{T}_n) \tag{34}$$

$$\hat{W}_\alpha(\mathbf{r}_j - \mathbf{T}_n) = \frac{1}{N_s} \sum_{\mathbf{k}} e^{-i\mathbf{k}\cdot\mathbf{T}_n} \hat{\psi}_{\mathbf{k},\alpha}(\mathbf{r}_j). \tag{35}$$

These site-symmetry adapted wannier functions and the Bloch functions transform as

$$\hat{g}\hat{W}_\alpha(\mathbf{r}_j - \mathbf{T}_n) = \sum_{\alpha'=1}^{n_\alpha} d_{\alpha'\alpha}(R_{j'}^{-1} R R_j) W_{\alpha'j'}(\mathbf{r} - \mathbf{T}_{j'j} - R\mathbf{T}_n) \tag{36}$$

$$\hat{g}\Psi_{\mathbf{k},\alpha}(\mathbf{r}_j) = \sum_{\alpha'=1}^{n_\alpha} d_{\alpha'\alpha}(R_{j'}^{-1} R R_j) \psi_{R\mathbf{k},\alpha'j'}(\mathbf{r}) e^{-iR\mathbf{k}\cdot\mathbf{T}_{j'j}}, \tag{37}$$

where $R = R_{j'} R_s R_j^{-1}$ and $\mathbf{T}_{j'j} = g\mathbf{R}_j - \mathbf{R}_{j'}$. The index $j'$ is determined by $g$ and $g_j$ to maintain $\mathbf{T}_{j'j}$ being lattice vectors. Then the irreducible representation of $G$ is given by

$$D_{\alpha'j',\alpha j}(g, \mathbf{k}) = e^{-iR\mathbf{k}\cdot\mathbf{T}_{j'j}} d^{(\beta)}_{\alpha'\alpha}(R_{j'}^{-1} R R_j). \tag{38}$$

The unitary matrix (22) for the maximally localized Wannier functions with the site-symmetry adapted can be obtained by solving

$$U_{n\mu}(R\mathbf{k}) = \tilde{d}_{n,n'}(g, \mathbf{k}) U_{n',\mu'}(\mathbf{k}) D^\dagger_{\mu',\mu}(g, \mathbf{k}), \tag{39}$$

where

$$\tilde{d}_{n,n'}(g, \mathbf{k}) = \int d\mathbf{r}\, \hat{\psi}_{n,R\mathbf{k}}^\dagger(\mathbf{r})\hat{\psi}_{n',\mathbf{k}}(g^{-1}\mathbf{r}) \tag{40}$$

with $\hat{\psi}_{n\mathbf{k}}(\mathbf{r})$ the eigenfunctions of the Bloch Hamiltonian. Here, we used the notation, $\mu = \{\alpha j\}$. With these $D_{\alpha'j',\alpha j}(g, \mathbf{k})$, $d_{\alpha'\alpha}(R_s)$, and $\tilde{d}_{n',n}(g, \mathbf{k})$ as inputs for the Wannier90 package [10], we can find the site-symmetry adapted Bloch wavefunction, Eq. (34), as

$$\hat{\Psi}_{\mathbf{k},\alpha}(\mathbf{r}_j) = \sum_n U_{n,\alpha j}(\mathbf{k})\, \hat{\psi}_{n\mathbf{k}}(\mathbf{r}), \tag{41}$$

leading to the site-symmetry enforced Wannier functions, Eq. (35).

## TIGHT-BINDING HAMILTONIAN

### Hopping amplitude

The most general tight-binding Hamiltonian is of the form

$$\mathcal{H}_0 = \sum_{\alpha\beta=\pm} \sum_{j,j'=BA,AB} t_{\alpha\beta}(\mathbf{R}_j - \mathbf{R}_{j'})\, d_\alpha^\dagger(\mathbf{R}_j)\, d_\beta(\mathbf{R}_{j'}), \tag{42}$$

where the creation operator $d_\alpha(\mathbf{R}_j)$ is associated with the Wannier functions

$$|\mathbf{R}_j, \alpha\rangle = d_\alpha^\dagger(\mathbf{R}_j)|0\rangle. \tag{43}$$

Then the hopping amplitude $t_{\alpha\beta}(\mathbf{R}_{jj'})$ is given by

$$\begin{aligned}
t_{\alpha\beta}(\mathbf{R}_{jj'}) &= \langle \mathbf{R}_j, \alpha|\, H\, |\mathbf{R}_{j'}, \beta\rangle \\
&= \int d\mathbf{r} d\mathbf{r}'\, \hat{W}_\alpha^\dagger(\mathbf{r}_j)\, H(\mathbf{r}, \mathbf{r}')\, \hat{W}_\beta(\mathbf{r}'_{j'}) \\
&= \frac{1}{N_s} \sum_{\mathbf{k},\alpha'} U_{\alpha'\alpha}^*(\mathbf{k})U_{\alpha'\beta}(\mathbf{k})\, e^{-i\mathbf{k}\cdot\mathbf{T}_{jj'}} \int d\mathbf{r}d\mathbf{r}'\, \hat{\psi}_{\alpha'\mathbf{k}}^\dagger(\mathbf{r})H_\mathbf{k}(\mathbf{r}, \mathbf{r}')\, \hat{\psi}_{\alpha'\mathbf{k}}(\mathbf{r}') \\
&= \frac{1}{N_s} \sum_{\mathbf{k},\alpha'} U_{\alpha'\alpha}^*(\mathbf{k})U_{\alpha'\beta}(\mathbf{k})\, \varepsilon_{\alpha'}(\mathbf{k})\, e^{-i\mathbf{k}\cdot\mathbf{T}_{jj'}},
\end{aligned} \tag{44}$$

where $\mathbf{T}_{jj'} = \mathbf{T}_{j'} - \mathbf{T}_j$, with $\mathbf{T}_j$ the lattice vector of the cell enclosing $\mathbf{R}_j$.

### Interaction Hamiltonian

The exchange interaction follows from the Coulomb interaction

$$\mathcal{H}_C = \frac{1}{2} \int d\mathbf{r}d\mathbf{r}'\rho(\mathbf{r})\frac{e^2}{|\mathbf{r} - \mathbf{r}'|}\rho(\mathbf{r}'), \tag{45}$$

where $\rho(\mathbf{r}) = \sum_\sigma \Psi_\sigma^\dagger(\mathbf{r}) \cdot \Psi_\sigma(\mathbf{r})$ are density operators written in terms of field operators

$$\Psi_\sigma(\mathbf{r}) = \sum_{\alpha,j} W(\mathbf{r} - \mathbf{R}_j)d_{j,\alpha,\sigma}.$$

The most general exchange term between sites $i, j$ that follows from (45) has the form

$$H_J = \sum_{i\neq j} \sum_{\alpha\beta\alpha'\beta'} \sum_{\sigma,\sigma'} J_{ij,\alpha\alpha'\beta\beta'}\, d_{i,\alpha,\sigma}^\dagger d_{j,\beta,\sigma'}^\dagger d_{i\beta',\sigma'}d_{j,\alpha',\sigma},$$

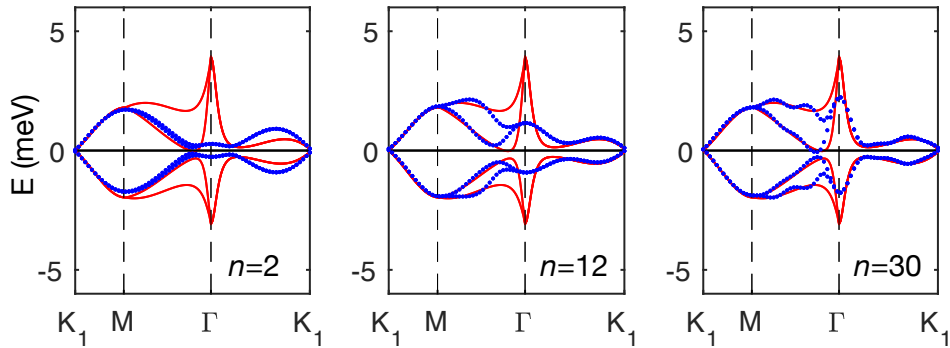

FIG. 1: Band structure constructed by the Wannier orbitals. The red solid lines represents the microscopic Bloch bands and the blue dots represent the Bloch bands constructed using the Wannier orbitals of the $n$ nearest-neighbors. $n = 2, 12$, and 30 correspond to the ranges $L = |\mathbf{T}|, 3|\mathbf{T}|$, and $5|\mathbf{T}|$, respectively. As $L$ increases, the values of $\Gamma$ point become closer to the microscopic Bloch bands.

where

$$J_{ij,\alpha\alpha'\beta\beta'} = \frac{1}{2}\int d^2r d^2r' \sum_{\alpha\alpha'\beta\beta'} W_\alpha^\dagger(\mathbf{r}-\mathbf{R}_i)\cdot W_{\alpha'}(\mathbf{r}-\mathbf{R}_j)\frac{e^2}{|\mathbf{r}-\mathbf{r}'|}W_\beta^\dagger(\mathbf{r}'-\mathbf{R}_j)\cdot W_{\beta'}(\mathbf{r}-\mathbf{R}_i) \quad (46)$$

is the exchange coupling. We find numerically that only the $J_{ij,\alpha\alpha\beta\beta} \equiv J_{ij,\alpha\beta}$ combination is non-zero. The exchange interaction has the form

$$\mathcal{H}_J = \sum_{i\neq j}\sum_{\alpha\beta}\sum_{\sigma,\sigma'} J_{ij,\alpha\beta}\, d_{i,\alpha,\sigma}^\dagger d_{j,\beta,\sigma'}^\dagger d_{i\beta,\sigma'}d_{j,\alpha,\sigma}. \quad (47)$$

Now mapping the $d$ operators into spin and pseudo-spin operators [9, 11],

$$d_{\alpha\sigma}^\dagger d_{\alpha\sigma} \to \left(\frac{1}{2}+\alpha\tau^z\right)\left(\frac{1}{2}+\sigma S^z\right)$$

$$d_{\alpha\sigma}^\dagger d_{\alpha-\sigma} \to \left(\frac{1}{2}+\alpha\tau^z\right) S^\sigma$$

$$d_{\alpha\sigma}^\dagger d_{-\alpha\sigma} \to \tau^\alpha\left(\frac{1}{2}+\sigma S^z\right)$$

$$d_{\alpha\sigma}^\dagger d_{-\alpha-\sigma} \to \tau^\alpha S^\sigma$$

with $\alpha = \pm$ and $\sigma = \pm$ and $\tau^\alpha = \tau^x + i\alpha\tau^y$, $S^\sigma = S^x + i\sigma S^y$, (47) can be cast into the form

$$\mathcal{H}_e = -\sum_{\sigma\sigma'}\sum_{\alpha\beta} J_{ij,\alpha\alpha}\left(\frac{1}{2}+2\tau_i^z\tau_j^z - 2\tau_i^x\tau_j^x - 2\tau_i^y\tau_j^y\right)\left(\frac{1}{2}+2\mathbf{S}_i\cdot\mathbf{S}_j\right). \quad (48)$$

## NUMERICAL DETAILS

Our calculations are carried out with the tight-binding Bloch Hamiltonian of the microscopic twisted graphene bilayer. The maximally localized Wannier orbitals were found with the Wannier90 package. The momentum space $\mathbf{k}$ mesh points were generated by the reciprocal supercell lattice vectors $\mathbf{G}_1$ and $\mathbf{G}_2$ by $\mathbf{k} = k_1\mathbf{G}_1 + k_2\mathbf{G}_2$ with $300\times 300$ grid points using periodic boundary conditions. This way, the Brillouin zone includes $\Gamma$, $K_1$, and $K_2$ high symmetric points. We modified the code in the package for the Brillouin zone to include all the $\mathbf{k}$ points and the corresponding $-\mathbf{k}$ points (modulo a reciprocal lattice vectors), reflecting the time-reversal conjugates of the Bloch wavefunctions.

In order to implement the Wannier90 package to find the maximally localized Wannier functions, we provided initial wavefunctions using the Bloch wavefunctions at the $\Gamma$ point as input functions. At that point, the Bloch wavefunctions

1st nearest-neighbor hoppings

| $(m,n)$ | $(0,0)$ | $(0,-1)$ | $(-1,0)$ |
|---|---|---|---|
| $t_{++}^{(BA,AB)}(\boldsymbol{\delta})$ | $0.004 + 0.385i$ | $0.004 + 0.385i$ | $0.004 + 0.385i$ |
| $t_{--}^{(BA,AB)}(\boldsymbol{\delta})$ | $0.121 - 0.363i$ | $0.121 - 0.363i$ | $0.121 - 0.363i$ |

2nd nearest-neighbor hoppings

| $(m,n)$ | $(0,1)$ | $(0,-1)$ | $(-1,0)$ | $(1,0)$ | $(-1,1)$ | $(1,-1)$ |
|---|---|---|---|---|---|---|
| $t_{++}^{(BA,BA)}(\boldsymbol{\delta})$ | $0.004 + 0.002i$ | $0.004 - 0.002i$ | $-0.000 + 0.000i$ | $-0.000 - 0.000i$ | $0.004 - 0.002i$ | $0.004 + 0.002i$ |
| $t_{++}^{(AB,AB)}(\boldsymbol{\delta})$ | $-0.001 - 0.002i$ | $-0.001 + 0.002i$ | $0.002 - 0.000i$ | $0.002 + 0.000i$ | $-0.002 + 0.002i$ | $-0.002 - 0.002i$ |
| $t_{--}^{(BA,BA)}(\boldsymbol{\delta})$ | $0.002 - 0.007i$ | $0.002 + 0.007i$ | $-0.002 - 0.009i$ | $-0.002 + 0.009i$ | $0.002 + 0.008i$ | $0.002 - 0.008i$ |
| $t_{--}^{(AB,AB)}(\boldsymbol{\delta})$ | $0.000 + 0.007i$ | $0.000 - 0.007i$ | $0.004 + 0.009i$ | $0.004 - 0.009i$ | $0.000 - 0.008i$ | $0.000 + 0.008i$ |

3rd nearest-neighbor hoppings

| $(m,n)$ | $(-1,-1)$ | $(-1,1)$ | $(1,-1)$ |
|---|---|---|---|
| $t_{++}^{(BA,AB)}(\boldsymbol{\delta})$ | $0.065 - 0.442i$ | $0.065 - 0.442i$ | $0.065 - 0.442i$ |
| $t_{--}^{(BA,AB)}(\boldsymbol{\delta})$ | $-0.071 + 0.441i$ | $-0.071 + 0.441i$ | $-0.071 + 0.441i$ |

4th nearest-neighbor hoppings

| $(m,n)$ | $(0,1)$ | $(0,-2)$ | $(1,0)$ | $(1,-2)$ | $(-2,0)$ | $(-2,1)$ |
|---|---|---|---|---|---|---|
| $t_{++}^{(BA,BA)}(\boldsymbol{\delta})$ | $-0.040 - 0.157i$ | $-0.039 - 0.157i$ | $-0.039 - 0.158i$ | $-0.040 - 0.157i$ | $-0.040 - 0.157i$ | $-0.040 - 0.157i$ |
| $t_{++}^{(AB,AB)}(\boldsymbol{\delta})$ | $-0.040 + 0.157i$ | $-0.039 + 0.157i$ | $-0.040 + 0.158i$ | $-0.040 + 0.157i$ | $-0.040 + 0.157i$ | $-0.040 + 0.157i$ |
| $t_{--}^{(BA,BA)}(\boldsymbol{\delta})$ | $0.009 + 0.163i$ | $0.008 + 0.163i$ | $0.008 + 0.163i$ | $0.009 + 0.163i$ | $0.009 + 0.163i$ | $0.008 + 0.163i$ |
| $t_{--}^{(AB,AB)}(\boldsymbol{\delta})$ | $0.009 - 0.163i$ | $0.008 - 0.163i$ | $0.008 - 0.163i$ | $0.009 - 0.163i$ | $0.009 - 0.163i$ | $0.008 - 0.163i$ |

5th nearest-neighbor hoppings

| $(m,n)$ | $(1,1)$ | $(-1,-1)$ | $(-1,2)$ | $(1,-2)$ | $(-2,1)$ | $(2,-1)$ |
|---|---|---|---|---|---|---|
| $t_{++}^{(BA,BA)}(\boldsymbol{\delta})$ | $0.011 + 0.081i$ | $0.011 - 0.081i$ | $0.013 - 0.084i$ | $0.013 + 0.084i$ | $0.008 + 0.084i$ | $0.008 - 0.084i$ |
| $t_{++}^{(AB,AB)}(\boldsymbol{\delta})$ | $0.009 + 0.085i$ | $0.009 - 0.085i$ | $0.007 - 0.082i$ | $0.007 + 0.082i$ | $0.012 + 0.082i$ | $0.012 - 0.082i$ |
| $t_{--}^{(BA,BA)}(\boldsymbol{\delta})$ | $0.011 - 0.084i$ | $0.011 + 0.084i$ | $0.014 + 0.083i$ | $0.014 - 0.083i$ | $0.006 - 0.081i$ | $0.006 + 0.081i$ |
| $t_{--}^{(AB,AB)}(\boldsymbol{\delta})$ | $0.009 - 0.082i$ | $0.009 + 0.082i$ | $0.007 + 0.083i$ | $0.007 - 0.083i$ | $0.014 - 0.085i$ | $0.014 + 0.085i$ |

6th nearest-neighbor hoppings

| $(m,n)$ | $(0,2)$ | $(0,-2)$ | $(-2,0)$ | $(2,0)$ | $(-2,2)$ | $(2,-2)$ |
|---|---|---|---|---|---|---|
| $t_{++}^{(BA,BA)}(\boldsymbol{\delta})$ | $0.005 + 0.007i$ | $0.005 - 0.007i$ | $-0.005 + 0.001i$ | $-0.005 - 0.001i$ | $-0.000 - 0.001i$ | $-0.000 + 0.001i$ |
| $t_{++}^{(AB,AB)}(\boldsymbol{\delta})$ | $-0.003 - 0.007i$ | $-0.003 + 0.007i$ | $0.006 - 0.001i$ | $0.006 + 0.001i$ | $0.002 + 0.001i$ | $0.002 - 0.001i$ |
| $t_{--}^{(BA,BA)}(\boldsymbol{\delta})$ | $0.006 - 0.003i$ | $0.006 + 0.003i$ | $-0.006 - 0.012i$ | $-0.006 + 0.012i$ | $0.001 + 0.010i$ | $0.001 - 0.010i$ |
| $t_{--}^{(AB,AB)}(\boldsymbol{\delta})$ | $-0.004 + 0.003i$ | $-0.004 - 0.003i$ | $0.008 + 0.012i$ | $0.008 - 0.012i$ | $0.001 - 0.010i$ | $0.001 + 0.010i$ |

TABLE I: The nearest-neighbor vectors $\boldsymbol{\delta} = m\mathbf{T}_1 + n\mathbf{T}_2$, where $\mathbf{T}_j$ are the superlattice primitive vectors. The superscript of $t_{\alpha\beta}^{(X,Y)}(\boldsymbol{\delta})$ indicates the hopping from the sublattice $|X,\alpha\rangle$ to $|Y,\beta\rangle$ with $X,Y = BA, AB$. In the momentum space, the Bloch Hamiltonian can be expressed by $t_{\alpha\beta}^{(X,Y)}(\mathbf{k}) = \sum_{\boldsymbol{\delta}} t_{\alpha\beta}^{(X,Y)}(\boldsymbol{\delta}) \exp(i\mathbf{k}\cdot\boldsymbol{\delta})$.

satisfy

$$\hat{C}_3 \, \hat{\psi}_{1(3),\Gamma}(\mathbf{r}) = \epsilon \hat{\psi}_{1(3),\Gamma}(\mathbf{r})$$

and

$$\hat{C}_3 \, \hat{\psi}_{2(4),\Gamma}(\mathbf{r}) = \epsilon^* \hat{\psi}_{2(4),\Gamma}(\mathbf{r}),$$

with $\epsilon = \exp^{i2\pi/3}$. The Bloch wavefunctions also satisfy $\hat{C}'_{2y} \psi_{1(2),\Gamma}(\mathbf{r}) = \hat{\psi}_{3(4),\Gamma}(\mathbf{r})$ with $\psi_{3(4),\Gamma}(\mathbf{r})$ being the $C_3$ eigenstates with $\epsilon^*(\epsilon)$, respectively. The four Wannier state spinors were derived from the input wavefunctions

$$\hat{W}_1^{(0)}(\mathbf{r}) = [\psi_{1,\Gamma}(\mathbf{r}_{A_1}), \psi_{3,\Gamma}(\mathbf{r}_{B_1}), \psi_{3,\Gamma}(\mathbf{r}_{A_2}), \psi_{1,\Gamma}(\mathbf{r}_{B_2})]^T,$$




\begin{thebibliography}{10}
\bibitem{note2}At low $T\ll J$, the zero-field susceptibility exhibits
characteristic 2D deviation from the conventional Curie-Weiss law,
$T\chi\sim(T/J)^{3}\exp{(2cJ/T)}$, while the magnetization is zero
at any finite $T$.
\end{thebibliography}
\end{document}